\documentclass[aps,pra,twocolumn,superscriptaddress]{revtex4-1}

\usepackage{graphicx}
\usepackage{amssymb}
\usepackage{subfigure}
\usepackage{amsmath}
\usepackage{amsfonts}
\usepackage{bm}
\usepackage{color}
\usepackage[dvipsnames]{xcolor}
\usepackage{physics}
\usepackage{ulem}

\usepackage{dcolumn}

\usepackage{lmodern}                         
\usepackage[T1]{fontenc}                   
\usepackage{ae}   
\usepackage[utf8]{inputenc}    
\usepackage{xfrac}  

\usepackage{palatino}
\usepackage[bottom]{footmisc}

\newcommand{\e}{\textup{e}}

\begin{document}

\title{Experimental Study of the Generalized Jarzysnki's Fluctuation Relation Using Entangled Photons}

\author{P. H. Souto Ribeiro}
\email{p.h.s.ribeiro@ufsc.br}
\affiliation{Departamento de F\'{i}sica, Universidade Federal de Santa Catarina, CEP 88040-900, Florian\'{o}plis, SC, Brazil}
\author{T. H\"{a}ffner}
\affiliation{Departamento de F\'{i}sica, Universidade Federal de Santa Catarina, CEP 88040-900, Florian\'{o}plis, SC, Brazil}
\author{G. L. Zanin}
\affiliation{Departamento de F\'{i}sica, Universidade Federal de Santa Catarina, CEP 88040-900, Florian\'{o}plis, SC, Brazil}
\author{N. Rubiano da Silva}
\affiliation{Departamento de F\'{i}sica, Universidade Federal de Santa Catarina, CEP 88040-900, Florian\'{o}plis, SC, Brazil}
\author{R. Medeiros de Ara\'{u}jo}
\affiliation{Departamento de F\'{i}sica, Universidade Federal de Santa Catarina, CEP 88040-900, Florian\'{o}plis, SC, Brazil}
\author{W. C. S. Silva}
\affiliation{Departamento de F\'isica, Universidade Federal de Alagoas, Arapiraca, AL, 57309-005, Brazil}
\author{R. J. de Assis}
\affiliation{Institute of Physics, Federal University of Goi\'{a}s, 74690-900, Goi\^{a}nia, GO, Brazil}
\author{L. C. C\'{e}leri}
\email{lucas@chibebe.org}
\affiliation{Institute of Physics, Federal University of Goi\'{a}s, 74690-900, Goi\^{a}nia, GO, Brazil}
\affiliation{Department of Physical Chemistry, University of the Basque Country UPV/EHU, Apartado 644, E-48080 Bilbao, Spain}
\author{A. Forbes}
\email{Andrew.Forbes@wits.ac.za}
\affiliation{School of Physics, University of the Witwatersrand, Johannesburg, South Africa}

\begin{abstract}
Optical modes possessing orbital angular momentum constitute a very useful platform for experimental studies on the quantum limits of Thermodynamics. Here, we present experimental results for entangled photon pairs subjected to thin turbulence simulated with spatial light modulators and interpret them in the context of the generalized Jarzysnki's fluctuation relation. By holographic measurement of the orbital angular momentum, we obtain the work distribution produced by the turbulence for single and double-sided turbulence channels. The use of the Klyshko's advanced wave picture allows us to interpret the experimental scheme as two-way processes in a fully quantum picture.
\end{abstract}

\maketitle

\section{Introduction}

Light beams possessing orbital angular momentum (OAM) have become an important subject in optical sciences \cite{padgett2017orbital}. They are considered promising candidates for several applications, notably classical \cite{willner2015optical} and quantum \cite{forbes2019quantum} communication. Regarding Quantum Thermodynamics, it has been shown \cite{Araujo18} that OAM beams can be naturally employed to experimentally investigate Jarzynski's fluctuation relation \cite{jar1997}. The link between photonic OAM states and Thermodynamics is made through the analogy between the paraxial wave equation and the time-dependent Sch\"{o}dinger equation in two dimensions \cite{Marcuse,Zanin2019}.

The interest in quantum Thermodynamics has hugely increased in the last decade, specially due to experimental developments in controlling quantum systems, quantum computation and quantum communication. From the experimental point of view, investigations of Thermodynamics of quantum systems were reported in nuclear magnetic resonance \cite{Batalhao2014,celeri,Assis19}, superconducting devices \cite{Cottet17} and ion traps \cite{Yan18} and other platforms are being considered \cite{Binder18}. Recently, the OAM degree of freedom of light was employed in order to simulate the Thermodynamics behaviour of a two-dimensional quantum mechanical oscillator undergoing a given unitary process \cite{Araujo18}, thus establishing a new experimental platform. Among other features, this system lives in a high dimensional Hilbert space, and thus can be employed to increase the rate of information transmission through a quantum channel.

In this article we go far beyond Ref. \cite{Araujo18} by considering entangled photons (in the OAM degree of freedom) undergoing a turbulent process that mimics the atmospheric turbulence. As we show, this process is non-unital (process that do not preserve the identity) and the generalized form of Jarzynski equality derived by Rastegin and \.Zyczkowski must be employed \cite{Rastegin14}. The results reported here show that the OAM degree of freedom of light can be employed to the experimental investigation of the quantum limits of Thermodynamics in the deep quantum limit.

\section{Quantum Thermodynamics with light beams}

We start by defining the class of protocols we are interested in. Let a system, described by an initial Hamiltonian ${\cal H_I}$, be prepared in a thermal equilibrium state (Gibbs ensemble). The protocol starts with an energy measurement (a projective measurement performed on the energy eigenbasis of the initial Hamiltonian). After this, a given process is applied to the system by some external agent (for instance, by changing some parameter of the Hamiltonian). The process can change the state, the Hamiltonian, or both. At the end of the protocol, a second energy measurement is performed, as a projection onto the eigenbasis of the final Hamiltonian ${\cal H_F}$. This the so called two-measurement protocol \cite{jar1997}. Notice that even though one starts with a Gibbs state, the first energy measurement projects it onto an energy eigenstate, so that the process is applied to pure states.

In our experiment, we consider the orbital angular momentum of light as the system, which undergoes a turbulence process. The system is described by the two-dimensional quantum harmonic oscillator, whose Hamiltonian is given by
\begin{equation}
H = (N_r + N_l + 1)\hbar\omega,
\label{hamiltonian}
\end{equation}
where $N_{r(l)}$ is the number operator for right (left) circular quanta. The eigenfuctions of this Hamiltonian are the Laguerre-Gaussian modes, which are characterized by an azimuthal quantum number $\ell$ (the topological charge) and by the radial quantum number $p$. For $p=0$, such modes have a ring-like transversal intensity profile, whose radii increases with $\ell$ \cite{Allen99}, and a phase singularity at the center (along the propagation axis). Since the eigenenergies of our system are given by $\epsilon_\ell = (|\ell|+1)\hbar\omega$, projections onto the OAM basis are equivalent to projections in the energy eigenbasis. In the two-measurement protocol, in each run of the experiment, the system is initially prepared in a Gibbs state and the first measurement projects it onto some OAM value $\ell$. The process is applied to this pure OAM state, which may evolve to a supperposition or an incoherent mixture of OAM states. The second measurement projects it onto a pure OAM state $\ell'$. The work done per run of the experiment when the system is driven by the process from an initial $\ell$ to a final $\ell^\prime$, can be defined as $W_{\ell, \ell'} = |\ell'| - |\ell|$ \cite{Araujo18} (for simplicity, we write from now on all energies in units of $\hbar\omega$). Since the outcome of each measurement is described by a random variable, the associated work probability distribution is then defined as
\begin{equation}
P(W) = \sum_{\ell,\ell'} p_{\ell\ell'}\delta\left(W - W_{\ell \ell'}\right),
\label{pw}
\end{equation} 
where $p_{\ell\ell'} = p_{\ell}p_{\ell'|\ell}$ is the joint probability of observing $\ell$ in the first measurement and $\ell'$ in the second one. $p_{\ell}$ is the probability of measuring $\ell$ before the process while $p_{\ell'|\ell}$ is the conditional probability of obtaining $\ell'$ after the process given that we observed $\ell$ before it. In the context of Thermodynamics, $p_{\ell}$ is  given by a thermal distribution associated with  the so called Gibbs state:

\begin{equation}
 \rho_{\beta} = \frac{\e^{-\beta H}}{Z}, 
 \label{gibbs}
 \end{equation}
 where the partition function is given by $Z = \e^{-\beta}\coth\left(\beta/2\right) - 1$ (note that the whole spectrum, except for $\ell = 0$, is doubly degenerated), and  $\beta = 1/k_B T$ stands for the inverse temperature. $k_B$ is the Boltzmann constant and T is the absolute temperature. For the optical system in our set-up, the physical realizaton of the thermal state corresponds to a single photon mixed state, where the populations of the density matrix represented in the energy eigenbasis are given by: 
 \begin{equation}
p_{\ell} = \frac{\e^{-\beta (|\ell|+1)\hbar \omega}}{Z}.
 \label{pops}
 \end{equation}

It has been recently demonstrated that this state arises naturally for the heralded idler photon in spontaneous parametric down-conversion (SPDC) by detection of the signal photon and tracing over its OAM degrees of freedom \cite{Haffner20}. The transition probabilities $p_{\ell'|\ell}$ contain the information about how the process acts on each energy eigenmode. This is actually the quantity that we measure in the experiment. 

While the coefficients  $p_{\ell}$ do have a physical meaning in terms of the single photon thermal state as we just described, in our approach it will be calculated as a function of $\beta$. In this way, we will be able to present our results as a function of $\beta$. 

In this regime, where fluctuations are important, thermodynamic quantities like work, heat and entropy becomes stochastic variables, thus implying that the usual laws of Thermodynamics do not hold. However, when fluctuations are properly taken into account, stronger laws emerge, in the form of fluctuation relations such as the Jarzynski equality \cite{jar1997}
\begin{equation}
    \langle \e^{-\beta W}\rangle = \e^{-\beta \Delta F},
    \label{eq:jar}
\end{equation}
where
\begin{equation}
\langle \e^{-\beta W}\rangle = \int dW P(W)  \e^{-\beta W},
\label{average}
\end{equation}
and $\Delta F$ is the free energy difference between the final and initial states (after and before the process is applied). The ensemble average is taken over the probability distribution (\ref{pw}). The usual form of the second law of Thermodynamics, $\expval{W} - \Delta F \geq 0$, follows from the convexity of the exponential function. This relation clearly states the statistical nature of the second law, that must hold only on average, and not on each realization of the process. It is important to observe here that this equation is valid as long as the dynamics is unital, i.e., the dynamical map describing the process preserves the identity.

This result was latter extended for processes described by any completely positive and trace-preserving map, resulting in \cite{Rastegin14}
\begin{equation}
\expval{e^{\beta W}} = \e^{-\beta \Delta F}(1 + \delta),
\label{rastegin}
\end{equation}
where $\delta = \mbox{Tr}[\rho_{\beta}G_{\Phi}]$, with $G_{\Phi} = \Phi(\rho^{*}) - \rho^{*}$ ($\rho^{*}$ is the maximally mixed state), is a measure of how much the dynamical map $\Phi$ deviates from an unital one. For unital maps, $\Phi(\rho^{*})=\rho^{*}$, thus implying $\delta = 0$. Again, from the convexity of the exponential function we get the modified second law of Thermodynamics
\begin{equation}
\expval{W} - \Delta F \geq -\beta^{-1}\ln(1+\delta).
\label{2ndlaw}
\end{equation}

Our goal in this article is to investigate the action of a turbulent process on the photonic OAM from the perspective of these fluctuation relations. 

\section{Experiment}

\begin{figure}[h]
\includegraphics[width=\columnwidth]{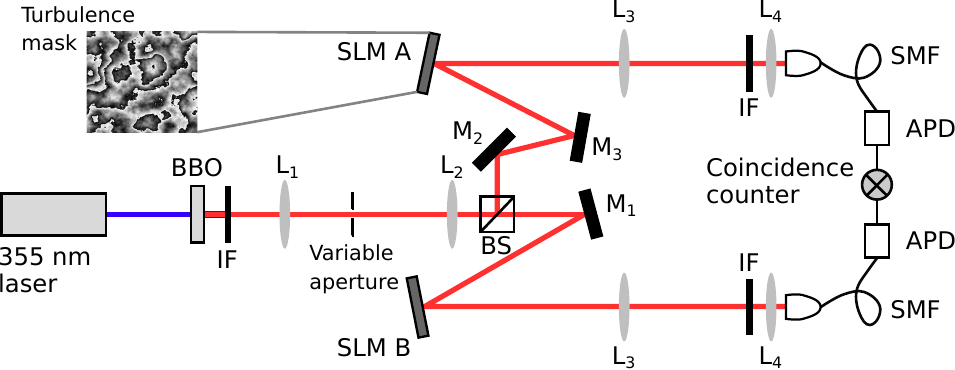}
\caption{Experimental setup. The plane of the crystal is imaged onto two separate SLMs using lenses L$_1$ and L$_2$ (f$_1$ = 200 mm and f$_2$ = 400 mm). Lenses L$_3$ and L$_4$ (f$_3$ = 500 mm and f$_4$ = 2 mm) were used to image the SLM planes, through 10-nm-bandwidth interference filters (IFs), to the inputs of the single-mode fibers (SMF). The photons coupled to the fibers were collected by avalanche photodiodes (APDs), which were further connected to a coincidence counter.}
\label{setup}
\end{figure}

\noindent Figure~\ref{setup} displays the experimental setup that was conceived to study the effect of atmospheric turbulence on quantum correlations between signal and idler photons, exploiting the OAM degree of freedom \cite{Hamadou13}. A 355-nm wavelength laser (mode-locked, average power of 350 mW) pumped a 3-mm-thick type-I $\beta$-barium borate (BBO) crystal, producing degenerate entangled photon pairs via spontaneous parametric down-conversion (SPDC). Each photon was sent through a 4$f$ telescope (lenses $L_1$ and $L_2$ with 200-mm and 400-mm focal distances, respectively), arranged to image the crystal plane onto the surfaces of two spatial light modulators (SLMs). The SLMs were used for generating and measuring the Laguerre-Gaussian modes, and for applying a turbulence process to one or both photons. Each SLM plane was imaged downstream using a second 4$f$ telescope (lenses $L_3$ and $L_4$ with 500-mm and 2-mm focal distances, respectively) onto an end of a single-mode optical fiber, into which only the fundamental Gaussian modes were coupled. The fibers terminate onto avalanche photodiodes (APDs), which were then connected to a photon coincidence counter. The coincidence counts registering the photon pairs were accumulated for 10 s with a gating time of 12 ns. Fluctuations on the pump beam caused an uncertainty in the measured counts of about 5\%.

For simulating the atmospheric turbulence as a one-sided or two-sided channel, random phase fluctuations were added to the phase mask of one or both SLMs. The fluctuations were distributed according to the Kolmogorov theory of turbulence \cite{Hamadou13}. Briefly, the random phase function is characterized by the scintillation strength $w_0/r_0$, where $w_0$ is the beam waist radius and $r_0$ is the Fried parameter \cite{Fried66}. Here, we used $w_0/r_0$ ranging from 0 to 4 with incremental steps of 0.2. The measurements for every scintillation strength were repeated 30 times, and the density matrix was reconstructed in each step via full quantum state tomography. After removing experimental imperfections leading to negative eigenvalues \cite{hamadou2012}, a mean density matrix for each scintillation strength was computed as the average of the reconstructed matrices. The effect of the turbulence on the orbital-angular-momentum entanglement was analysed elsewhere \cite{Hamadou13}. Here, we focus on investigating the work realized by the turbulence on the photonic OAM using the generalized Jarzynski's fluctuation relation. 


\section{Remote state preparation and the two-measurement protocol}

The quantum state of the entangled photons produced in the SPDC process can be written in a simplified form, in terms of an entangled state of the OAM degrees of freedom, as $\vert \psi \rangle = \sum_{\ell_{a}, \ell_{b}} C^{\ell_{a}, \ell_{b}}_{\ell_p} \,\, \vert \ell_{a}, \ell_{b} \rangle$, where $\ell_{p}$, $\ell_{a}$ and $\ell_{b}$ are the pump, signal and idler OAM azimuthal indices, respectively, while $C^{\ell_{a}, \ell_{b}}_{\ell_p}$ stands for the coefficients that take into account the phase matching function and the OAM of the pump beam \cite{spiral}. We will restrict ourselves to states where the radial index is $p=0$. 

\begin{figure}[h]
\includegraphics[width=\columnwidth]{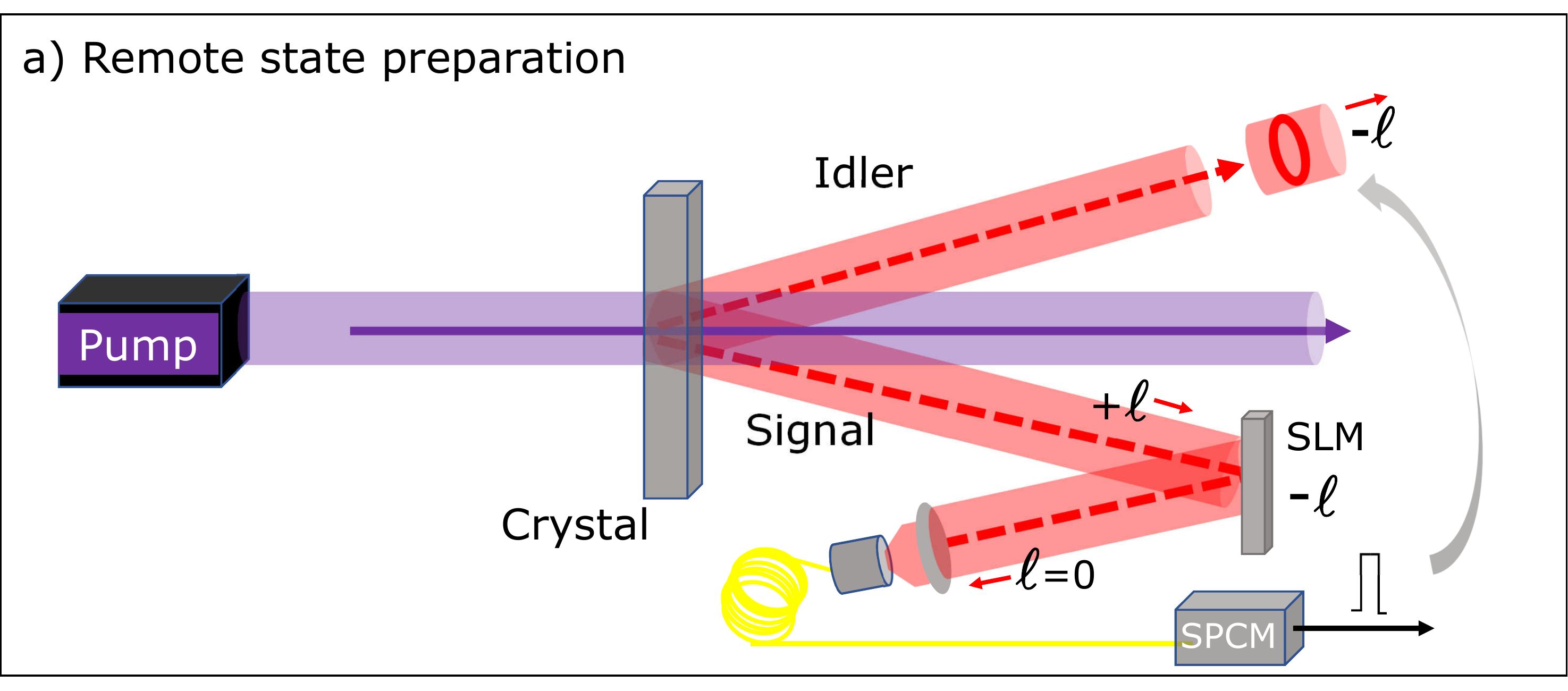}
\includegraphics[width=\columnwidth]{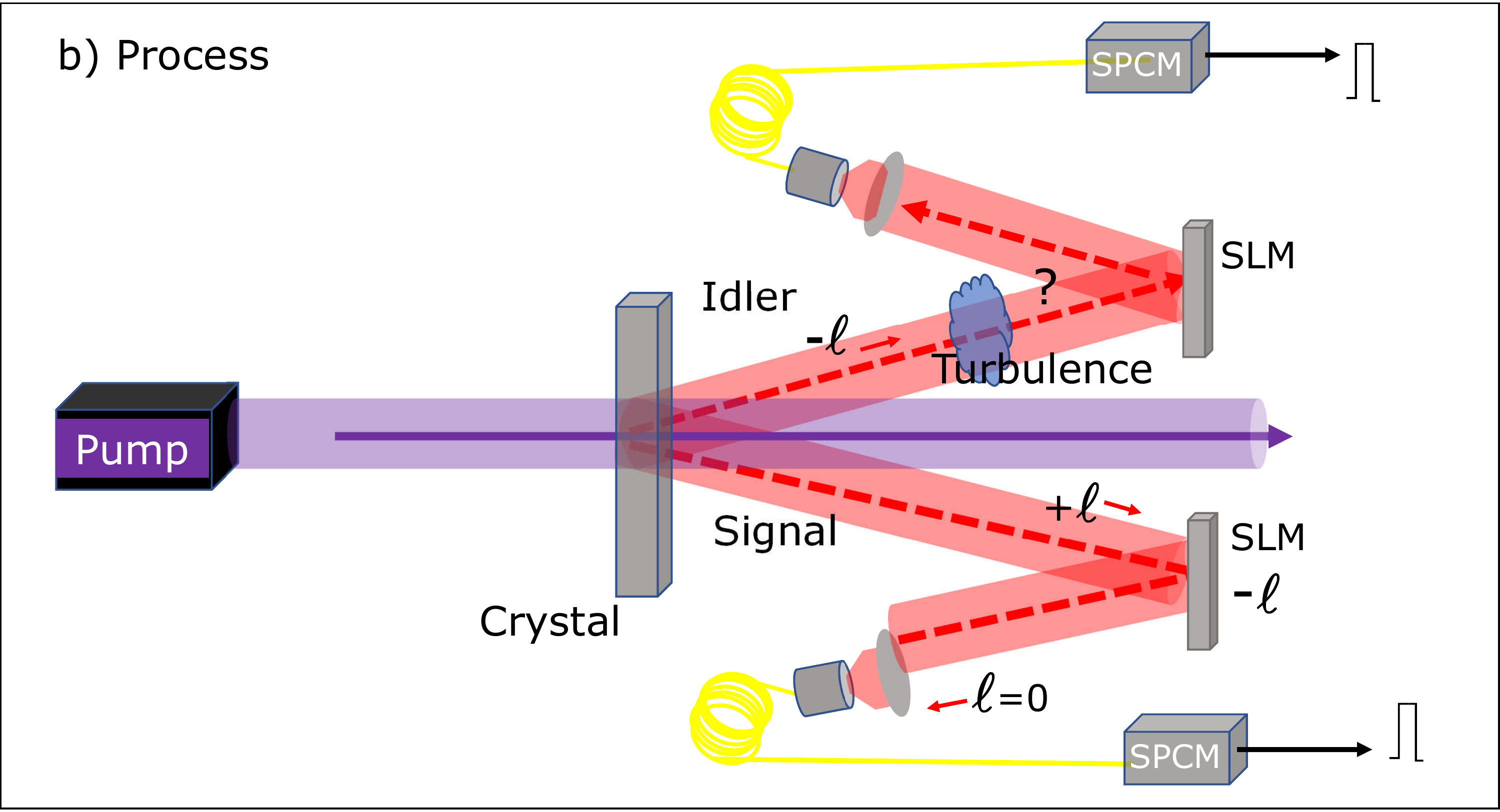}
\caption{a) Remote state preparation; b) Remote state preparation, process and second measurement.}
\label{remote}
\end{figure}

Considering the case where $\ell_p = 0$, the down-conversion state can be well approximated by \cite{spiral} 
\begin{equation}
\label{SPDC1}
\vert \psi \rangle = \sum_{\ell = -\infty}^{\infty} C_\ell  \,\, \ket{+\ell}_a\ket{-\ell}_b,
\end{equation}
where $C_{-\ell}=C_\ell\equiv C^{+\ell,-\ell}_0$. This is a high-dimensional entangled state where signal and idler modes have OAM with the same absolute value $\vert\ell\vert$, but with opposite signs. Let us consider now what is the state of the idler photon conditioned to a measurement performed on the signal, which  is  the  case of  our  experiment. In Fig. \ref{remote}a) this scenario is illustrated. In the case where the signal photon is detected, but its OAM is not measured (traced out), the idler photon is prepared in a heralded single photon thermal state \cite{Haffner20}. However, whenever the signal photon is detected and its OAM is measured to be $+\ell$, the idler photon is remotely prepared in a pure single photon state with  OAM = $-\ell$. Figure \ref{remote}a) illustrates the remote state preparation, while Fig. \ref{remote}b) illustrates the complete scheme: i) the idler photon is remotely prepared in the state with OAM = $-\ell$; ii) the idler photon propagates through the turbulence that acts on it as the process; iii) the process couples the idler photon with other OAM modes so that it evolves to a supperposition or a mixture of OAM states and SLM plus optical fiber projects it onto some final OAM.

The goal of this procedure is to measure the conditional probabilities $p_{\ell|\ell'}$ that a given initial OAM = $\ell$ (state) evolves into a final OAM = $\ell'$ (state) induced by the action of the turbulence. Notice that the transition probabilitiy $p_{\ell,\ell'} = p_{\ell}\,\, p_{\ell|\ell'}$ includes $p_{\ell}$ that depends on the temperature of the initial state and is calculated from Eq. (\ref{pops}).

\vspace{0.5cm}
\section{Results and discussion}

\begin{figure}
\centering
\includegraphics[width=\columnwidth]{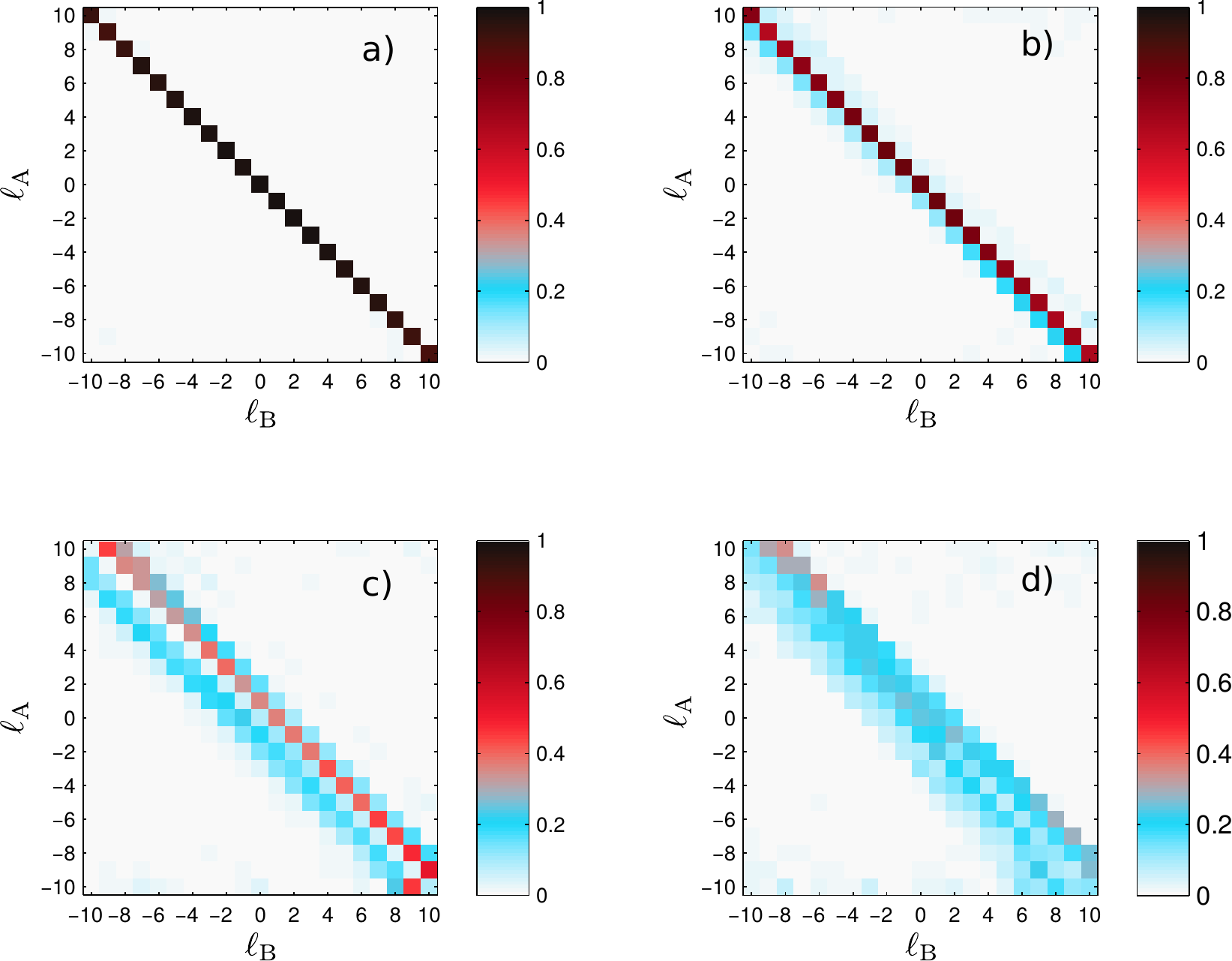}
\caption{ Conditional probability matrices of turbulence processes given by the coincidence counts for simultaneous measurements of modes with azimuthal index $\ell_A$ in the signal beam and $\ell_B$ in the idler beam when only one of the two photons (the idler) propagates through a) no turbulence, b)  mask 1, c)  mask 2 and d)  mask 4 turbulences with increasingly (from b) to c)) stronger scintillation strengths. }
\label{Matrix1}
\end{figure}

\begin{figure}
\centering
\includegraphics[width=\columnwidth]{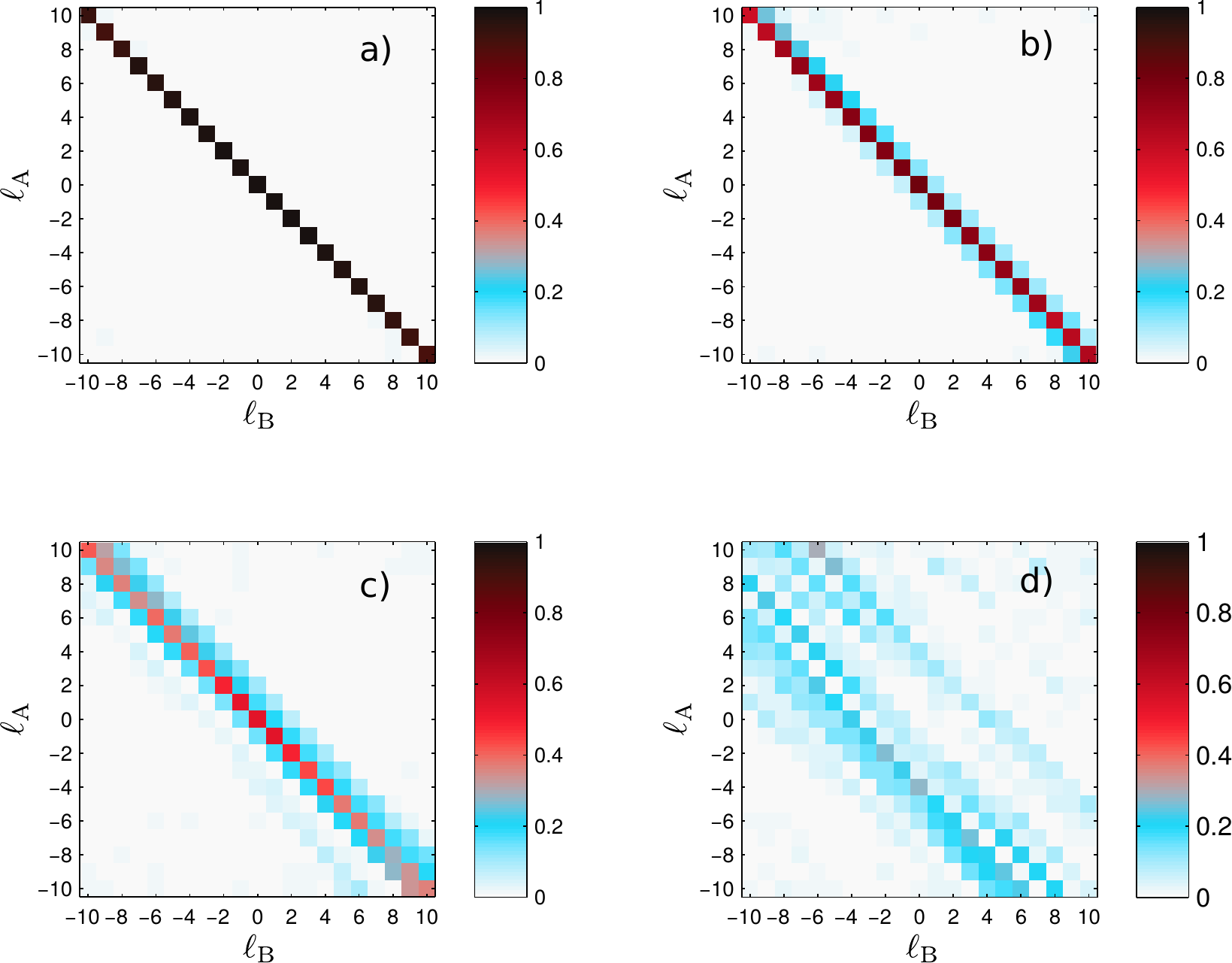}
\caption {Conditional probability matrices of turbulence processes given by the coincidence counts for simultaneous measurements of modes with azimuthal index $\ell_A$ in the signal beam and $\ell_B$ in the idler beam, when both photons propagate through a) no turbulence, b)  mask 1, c)  mask 2 and d)  mask 4 turbulences with increasingly (from b) to c)) stronger scintillation strengths.}
\label{Matrix2}
\end{figure}

\begin{figure}[htb]
\includegraphics[width=.9\columnwidth]{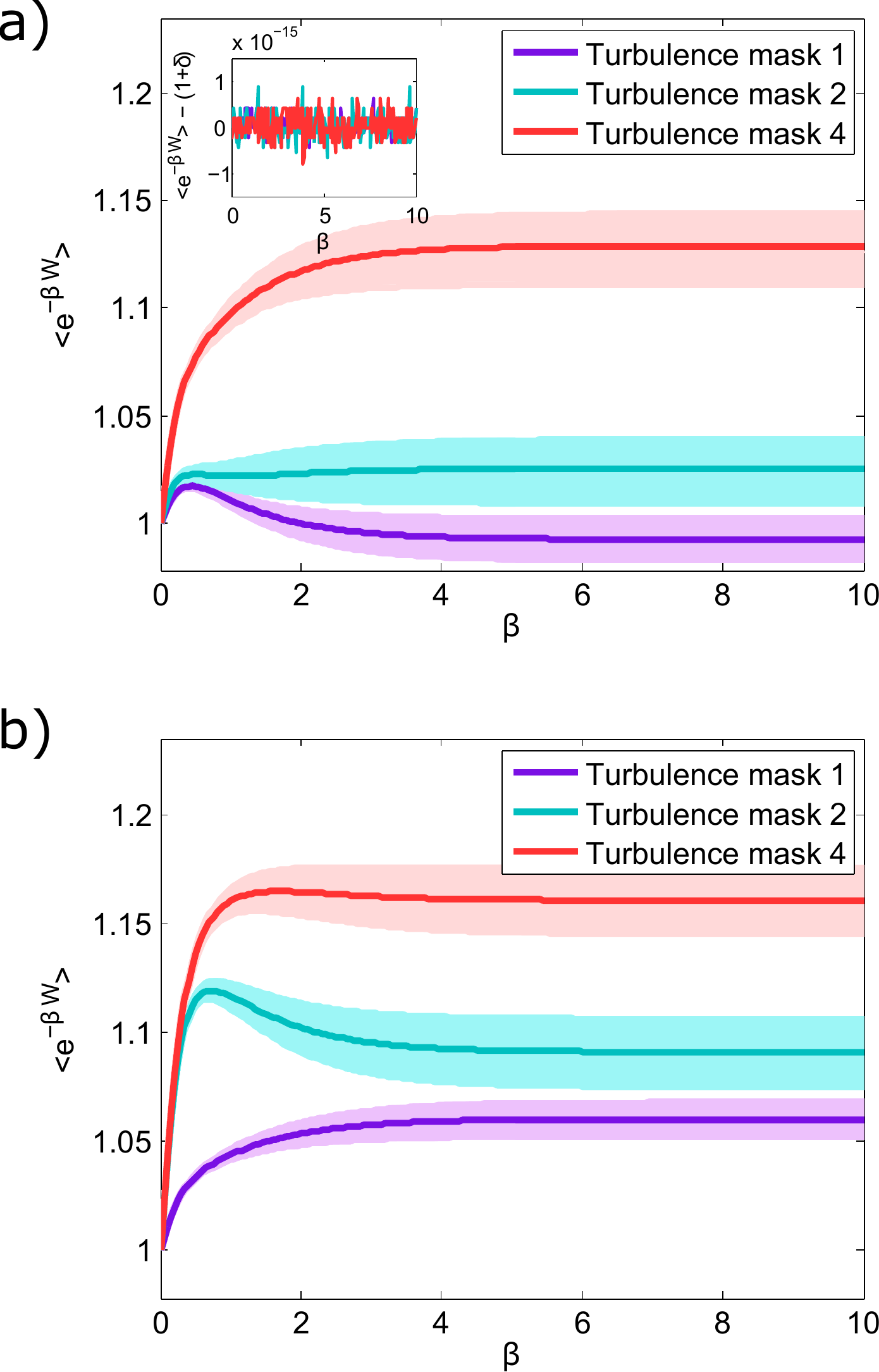}
\caption{$\left\langle e^{-\beta W}\right\rangle$ plotted as a function of $\beta$ for a) forward one-photon process and b) backward one-photon process. The inset in a) displays the deviation from the generalized Jarzynski fluctuation relation.}
\label{plots1}
\end{figure}

\begin{figure}[htb]
\includegraphics[width=.9\columnwidth]{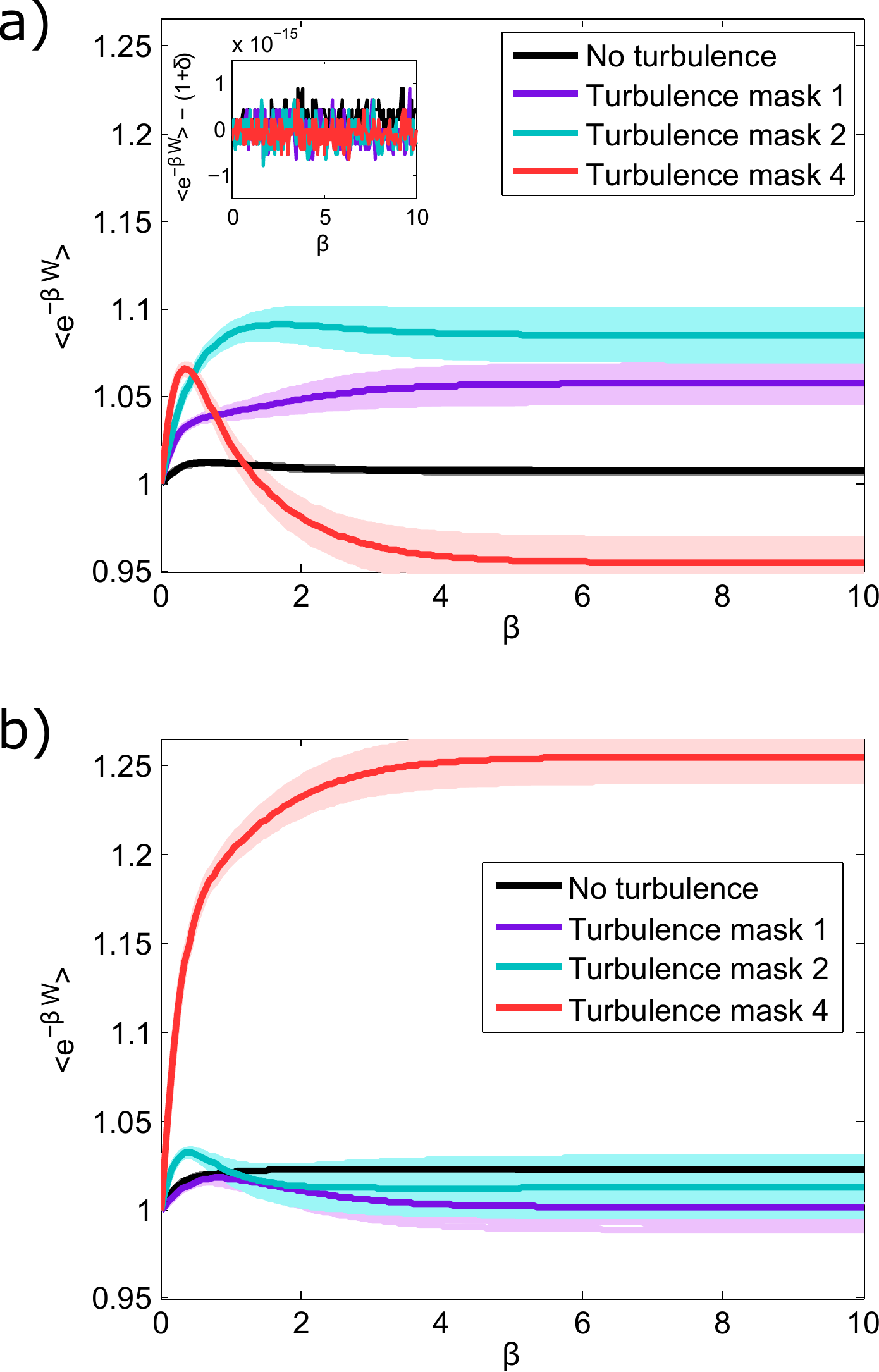}
\caption{$\left\langle e^{-\beta W}\right\rangle $ plotted as a function of $\beta$ for a) forward two-photon process and b) backward two-photon process. The inset in a) displays the deviation from the generalized Jarzynski fluctuation relation.}
\label{plots2}
\end{figure}

Measurement results are shown in Fig.~\ref{Matrix1} for the single-sided turbulence channel,  where only the idler photon is submitted to the turbulence process, and in Fig.~\ref{Matrix2} for double-sided channels, where both signal and idler photons are submitted to turbulence masks prepared in the SLMs. They provide the conditional transition probabilities $p_{\ell'|\ell}$. They are used to compute  the work probability distribution according to Eq.~\ref{pw}, and $P(W)$ is used to compute the quantity $\left\langle e^{-\beta W}\right\rangle $, with the average taken over $P(W)$. From the measurements shown in the absence of a turbulence process (Fig.~\ref{Matrix1}a or Fig.~\ref{Matrix2}a), we can  see that signal and idler have a high initial degree of correlation. This measurement results can be interpreted as a calibration procedure.  From the fluctuation relation perspective, the high quality calibration result can be interpreted as an experimental evidence that the free energy does not change. This interpretation is justified by the fact that the OAM modes are the physical realization of the Hamiltonian energy eigenstates. The nearly perfect correlation means that the remote state preparation is actually working as expected and also that the measurement procedure is performing projections onto the same initial state basis, or the same initial Hamiltonian eigenstates basis. Therefore,  $\Delta F = 0$, and one expects the quantity $\left\langle e^{-\beta W}\right\rangle $ to be equal to 1, in accordance to Jarzynski equality (Eq. \ref{eq:jar}).  We can also conclude from Figs. ~\ref{Matrix1} b), c) and d) and  Figs. ~\ref{Matrix2} b), c) and d) that the increasing scintilation strength results in an increasing loss of correlation, as expected.

 The results for $\left\langle e^{-\beta W}\right\rangle $ as a function of $\beta$ are shown in Fig.~\ref{plots1}a). They concern what we called forward process, which starts with the remote state preparation of the idler photon, application of the turbulence process on it and then measurement of the OAM. The results show that for the turbulence with higher scintilation strength (mask 4), there is a considerable deviation from Jarzynski's equality, while for the smaller scintilation strengths (mask 1 and mask 2) this deviation is small. The inset in Fig.~\ref{plots1}a) shows that the generalized Jarzynski's fluctuation relation is well respected for all turbulence masks. This is shown throught the deviation $\left\langle e^{-\beta W}\right\rangle - (1-\delta)$, which is nearly zero when $\Delta F$ = 0, according to Eq. \ref{rastegin}. We conclude that the turbulence acts as a non-unital process.
We recall that we are dealing with an infinite dimensional system, which is truncated in $\vert\ell\vert \leq 10$. For $\beta$ in the region below 2, the truncation effects are more pronounced and  may  introduce  non-physical  features,  like the peaks observed in a few of these curves, while for $\beta$ in the region above 2, they are negligible. For temperatures low enough (i.e., for $\beta$ high enough), the coefficients of the thermal distribution are very small for $\vert\ell\vert > 10$, making the contribution of high order modes negligible for Thermodynamics purposes. The shaded area represents the measurement uncertainties within 95\% confidence level, calculated by propagating the coincidence counting rate uncertainty using Monte Carlo processing.

 In Fig.~\ref{plots1}b) the results for the single-channel backward process are shown. It consists in the time-reversal of the forward process and it is physically interpreted in terms of the Klyshko's advanced wave picture (AWP) \cite{Klyshko88,Belinskii94} described in detail in the Appendix A.  We can se that there is also deviation from the  Jarzynski's equality. The results are also compatible with the generalized version of the fluctuation relation, according to the deviation $\left\langle e^{-\beta W}\right\rangle - (1-\delta) \simeq 0$, which has the same behaviour as in the case of the forward process (not shown). We would expect a symmetric behavior for forward and backward process. However, this is clearly not the case. 

 Figs.~\ref{plots2}a) and \ref{plots2}b) show the results for forward and backward processes respectively, for the case of turbulence applied to both signal and idler photons, also called double sided channel. The physical interpretation is also provided in terms of the AWP and the details are presented in Appendix A. This process is equivalent to propagation through a turbulent medium, followed by a free propagation and then through a second turbulent medium with the same scintilation strength. The results also follow the generalized Jarzinski's relation (see inset in Fig. \ref{plots2}a) similar to the inset in Fig. \ref{plots1}a) and are not symmetric. 

\begin{figure}[htb]
\centering
\includegraphics[width=.9\columnwidth]{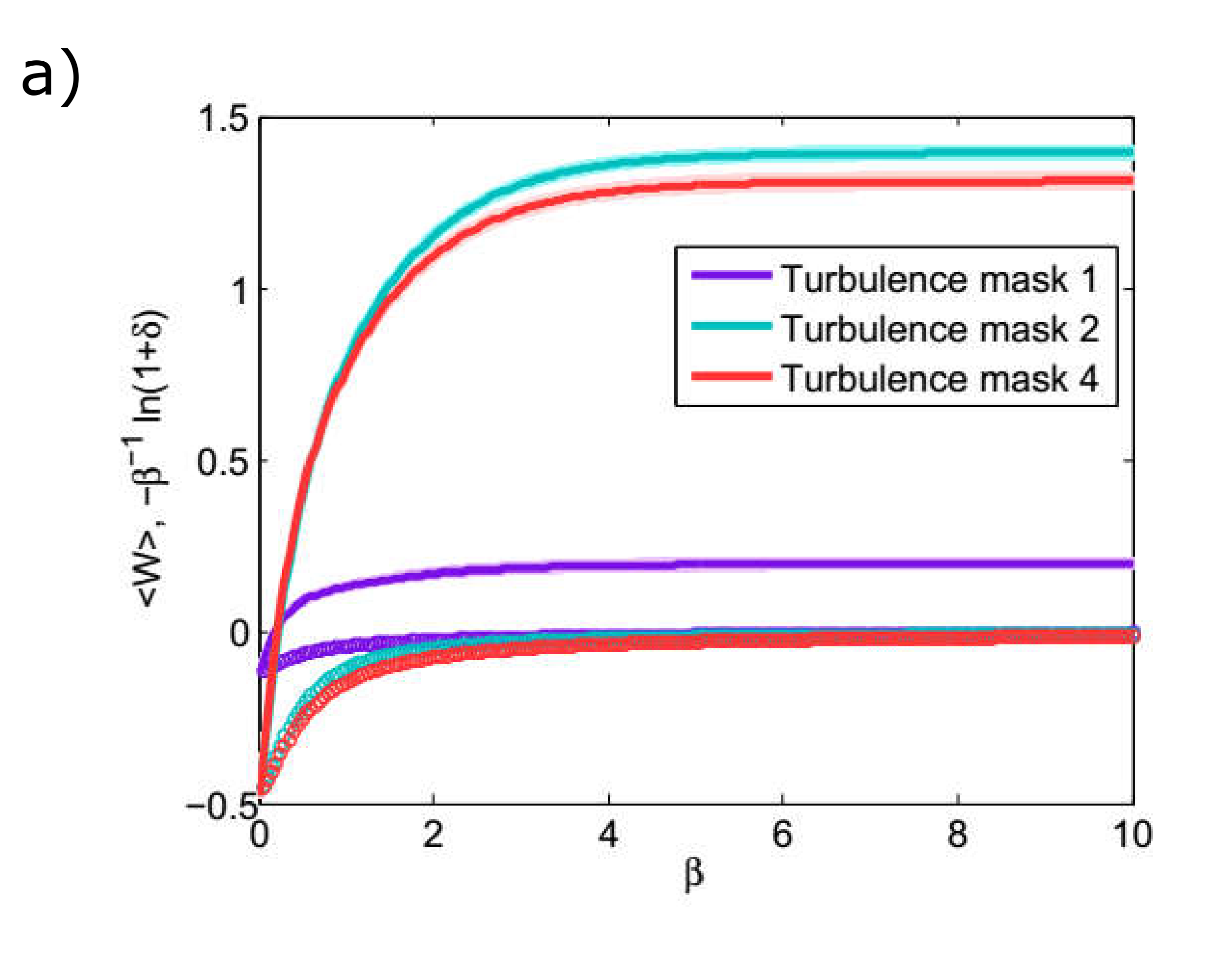}
\includegraphics[width=.9\columnwidth]{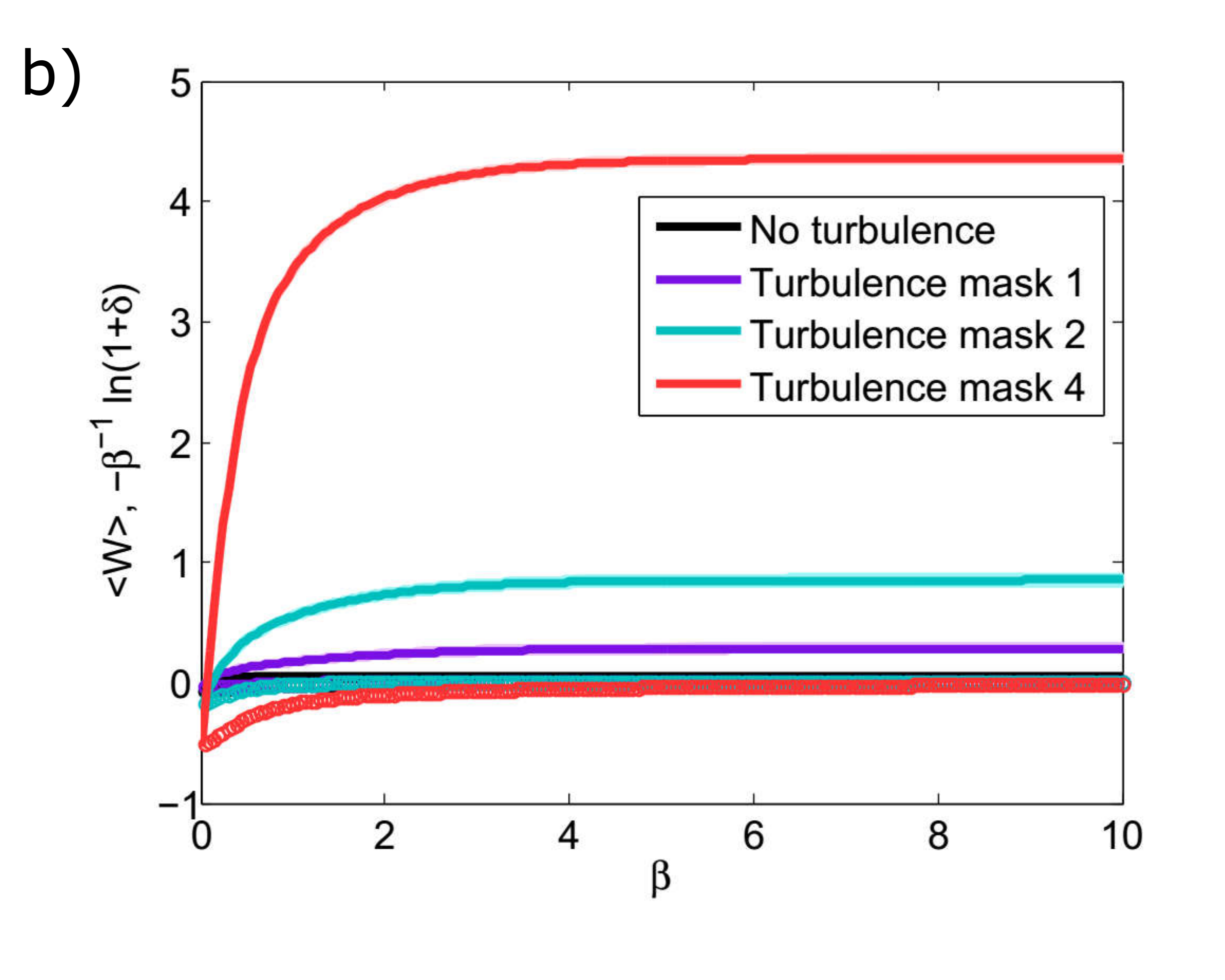}
\caption{$\left\langle  W \right\rangle$ and $-\beta^{-1}\ln(1+\delta)$ (circles) plotted as a function of $\beta$ for a) forward one-photon process and b) forward two-photon process.}
\label{plot2nd}
\end{figure}

We have also computed the average work $\langle W \rangle$  for the single sided and double sided forward channels, which respects the generalized form of the second law of Thermodynamics given in Eq. \ref{2ndlaw}. In Fig. \ref{plot2nd} we can see $\langle W \rangle$ and $-\beta^{-1}\ln(1+\delta)$ as a function of $\beta$. The error bars were also calculated using Monte Carlo simulation as before and are represented by the thickness of the curves. In all cases (backward processes not shown) $\left\langle  W \right\rangle$ is much bigger than $-\beta^{-1}\ln(1+\delta)$. Therefore, we can conclude that despite the non-unitality, which would allow  $\langle W \rangle < 0$ meaning that the light beam would realize work on the turbulence, the average work is always positive, meaning that the turbulence realizes work on the photons on average.  

\subsection{Non-unitality and loss of correlation}

We have used photon pairs entangled in their OAM degree of freedom to analyze the work realized by the Komolgorov turbulence on the photonic OAM. We concluded that it is a non-unital process that realizes work on the photons and that the average work tends to be bigger as the scintilation strength increases. We introduce the concept of forward and backward processes based on the Klysko's AWP and observe that the turbulence is not symmetric, meaning that propagations forward and backwards the turbulence are not equivalent.

The reason for both, non-unitality and asymmetry is a consequence of the method used to simulate atmospheric turbulence with SLMs. In order to emulate the dynamical effects of the turbulence, a given turbulence channel is implemented by applying SLM masks with the same scintilation strength that are periodically changed during the measurement time. While each (phase-only) SLM mask realizes a unitary transformation in the photonic field, the measurement results for several masks are summed uncoherently. This implies in a loss of phase information and makes the overall process irreversible and asymmetric. It is interesting to note that the fluctuation relation were able to detect this feature, in the conditions of our experiment. In Appendix B we provide an explanation about the effects of the loss of phase information and irreversibility in optical diffraction using the double-slit experiment as an example.

The measurement results presented in Figs.~\ref{Matrix1} and ~\ref{Matrix2} show the loss of corrrelation between signal and idler OAM in SPDC. In the optical communication community, the coupling between an OAM mode incident in a turbulent medium and other OAM modes at the output is called crosstalk. This term suggests that the turbulent medium allows the energy exchange betweem input and output OAM modes. It is intuitive that bigger scintilation strengths produce stronger crosstalks. One could try to define a parameter to quantify the crosstalk strength by counting the number of events where an input OAM = $\ell$ transfers its photon to an output OAM = $\ell'$. Noticing that some work must be done in order to increase or decrease the orbital angular momentum, we realize that the statistical distribution of thermodynamic work $P(W)$ is closely related to the amount of crosstalk. However, neither the crosstalk or $P(W)$ quantify the loss of correlation. For instance, we can have a process (not a turbulence) that simply increases the OAM by a fixed amount $+m$.  There will be thermodynamic work and there will be crosstalk, however the correlation will remain unaltered.


\section{Conclusion}

We analyse an experiment that measures the coupling between input and output OAM modes crossing a turbulent medium simulated by SLMs applying Komolgorov turbulence masks  from the point of view of quantum Thermodynamics via fluctuation relations. Specifically, we investigate the generalized Jarzynski's fluctuation relation. We interpret the scheme in terms of remote state preparation and Klyshko's advanced-wave picture (AWP), we observe that  this process is not unital due to the method to generate the dynamics of the turbulence and it was necessary to use the generalized Jarzynski's equality for quantum stochastic maps to fit the experimental data. For stronger scintillation masks and for double-sided channels, the deviation from unital processes was more significant. We have also analyzed the second law of Thermodynamics in terms of the average value of the work $\langle W \rangle$ and observed that, while the fluctuation relation allows $\langle W \rangle < 0$ for the non unital process, $\langle W \rangle $ is always grater than zero for our measurements. In conclusion, we have presented a new experimental scheme that allows the investigation of optical processes from the perspective of quantum fluctuation relations and stablish a promising tool for future investigations in the field.

\begin{acknowledgements}
The authors would like to thank the Brazilian Agencies CNPq, FAPESC, and the Brazilian National Institute of Science and Technology of Quantum Information (INCT/IQ). This study was financed in part by the Coordenação de Aperfeiçoamento de Pessoal de Nível Superior - Brasil (CAPES) - Finance Code 001. LCC would like to also acknowledge support from Spanish MCIU/AEI/FEDER (PGC2018-095113-B-I00), Basque Government IT986-16, the projects QMiCS (820505) and OpenSuperQ (820363) of the EU Flagship on Quantum Technologies and the EU FET Open Grant Quromorphic and the U.S. Department of Energy, Office of Science, Office of Advanced Scientific Computing Research (ASCR) quantum algorithm teams program, under field work proposal number ERKJ333.
\end{acknowledgements}
\vspace{0.5cm}
\appendix 

{\bf APPENDIX A}  \\  \\  {\it Klyshko's Advanced-Wave Picture}\\ 

\begin{figure}[h]
\includegraphics[width=\columnwidth]{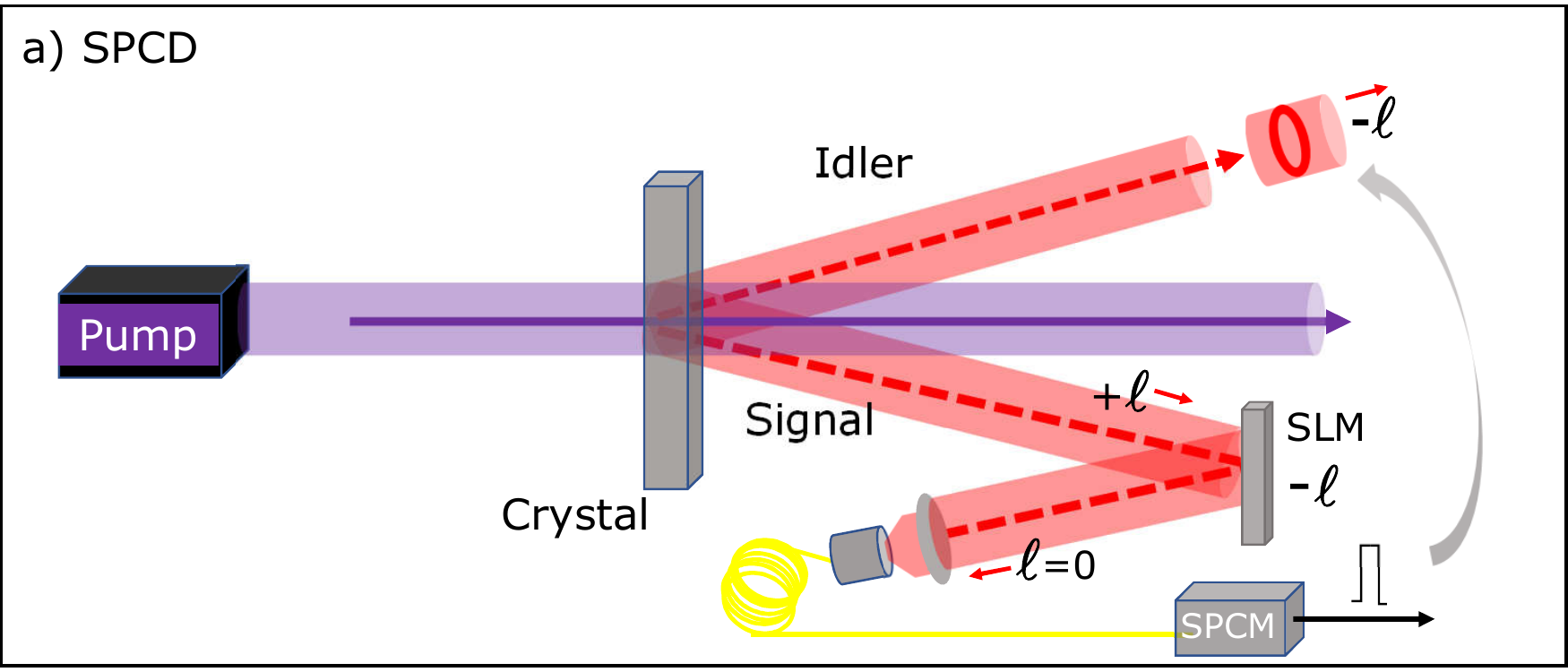}
\includegraphics[width=\columnwidth]{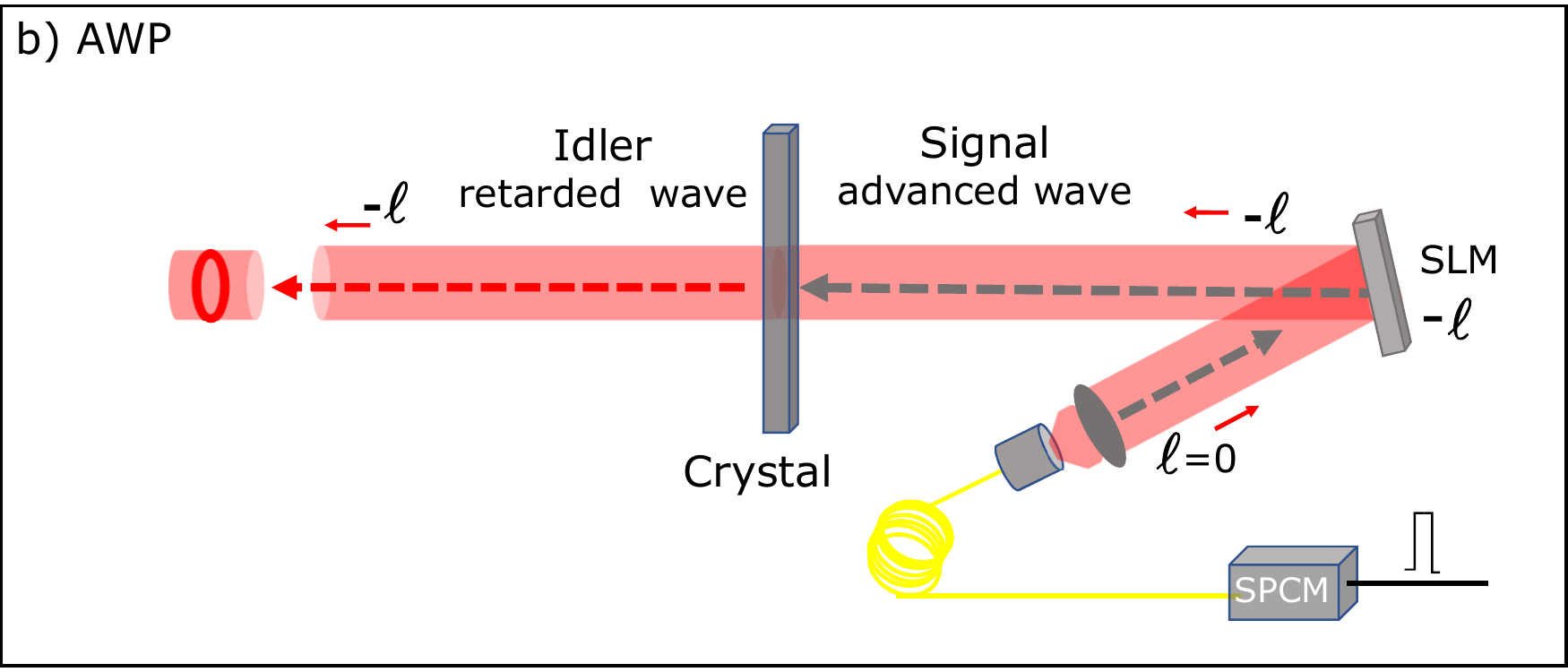}
\includegraphics[width=\columnwidth]{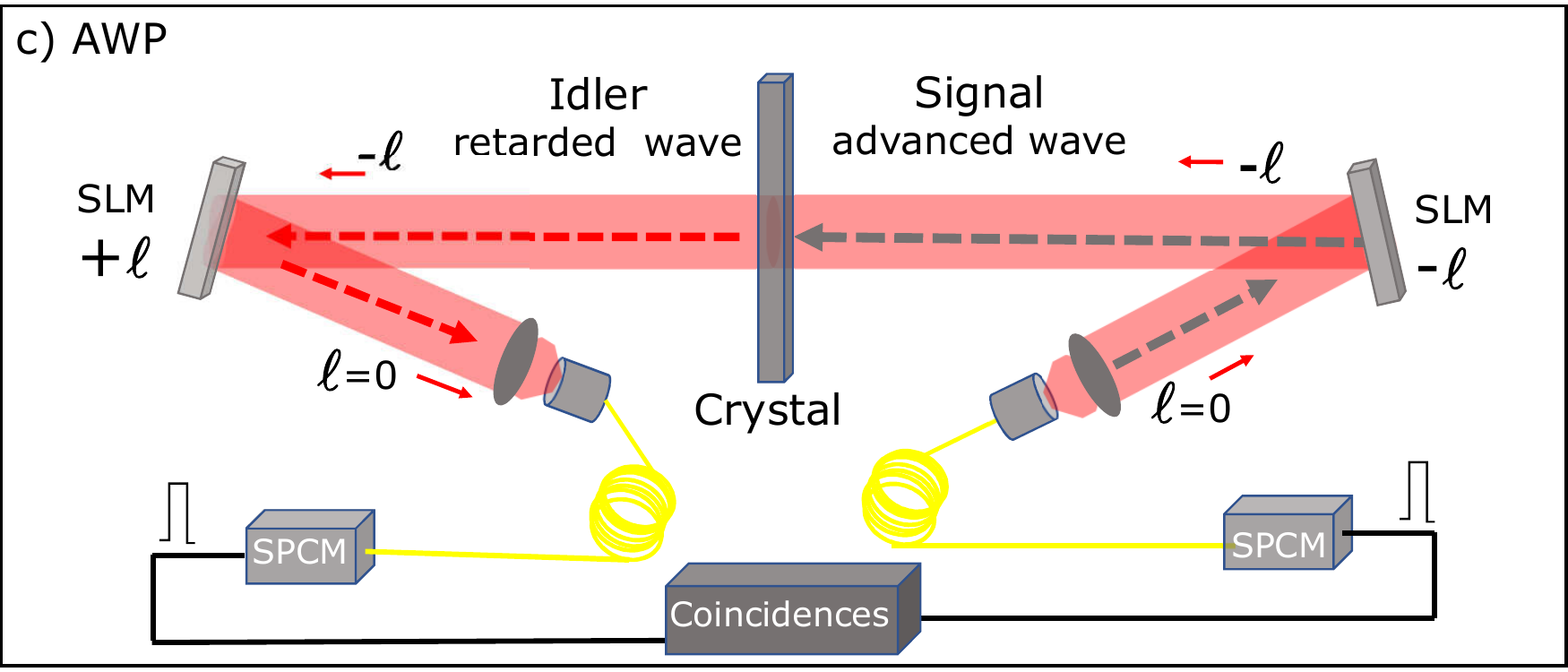}
\includegraphics[width=\columnwidth]{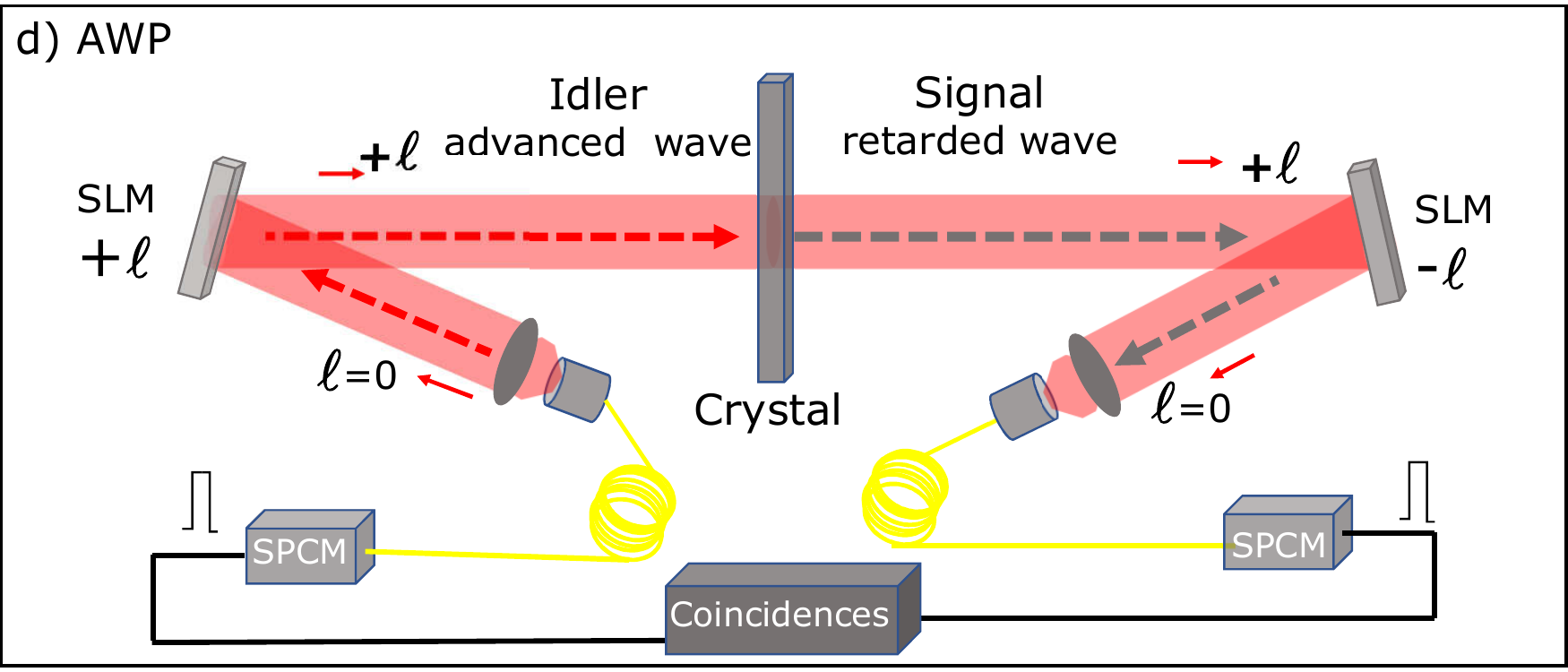}
\caption{Remote state preparation and Klyshko's advanced-wave picture (AWP). a) Remote state preparation. b) AWP of a). c) Direct AWP of OAM detection scheme. d) Reverse AWP of OAM detection scheme. See main text for details.}
\label{AWP1}
\end{figure}

The remote state preparation scenario can be understood in terms of Klyshko's {\it Advanced-Wave Picture} (AWP) \cite{Klyshko88,Belinskii94}. The AWP is a method that makes use of the quantum correlations between the transverse spatial degrees of freedom of signal and idler beams to provide a classical analog of the experimental picture \cite{Walborn10a}. Figure~\ref{AWP1}b) illustrates the AWP for the remote state preparation scheme shown in Fig.~\ref{AWP1}a). The SPCM detector acts like a light source emitting an advanced photonic wave packet that propagates backwards in the optical fiber and that is outcoupled and collimated by the objective. At this point, it is a zero-order Gaussian mode that propagates back towards the crystal and then forward to the idler detector. In its way back to the crystal it finds the SLM, which applies an operation $\cal{L}_{-}^{\ell}$ preparing the advanced beam with OAM~=~$-\hbar\ell$ per photon. The crystal acts like a SLM (transmission in this picture) that is controlled by the angular spectrum of the pump beam (not shown in this picture). In the present scheme, it acts as a transparent plate coupling the signal advanced wave with the idler retarded wave, which is thus prepared with OAM~=~$-\hbar\ell$ per photon. 
 
Figure~\ref{AWP1}c) is an extension of Fig.~\ref{AWP1}b) with the addition of an OAM detection scheme on the idler side. It indicates that the application of the operation $\cal{L}_{+}^{\ell}$ should bring the idler OAM to zero, then coupling to the optical fiber with high success rate. It is worth mentioning that the AWP works only for fields generating coincidence counting events, that is to say, for twin photon pairs. This implies that the AWP actually does not impose any particular direction for creating the {\it picture}, meaning that changing the roles of signal and idler beams as advanced and retarded waves (see Fig. \ref{AWP1}d)) should give rise to the same photon counting rate. Therefore, the scheme shown in Fig.~\ref{AWP1}d is equivalent to that of Fig.~\ref{AWP1}c). However, in this case it is the idler SPCM detector that is interpreted as the light source, while the signal SPCM is viewed as the actual detector.

\begin{figure}[h]
\includegraphics[width=\columnwidth]{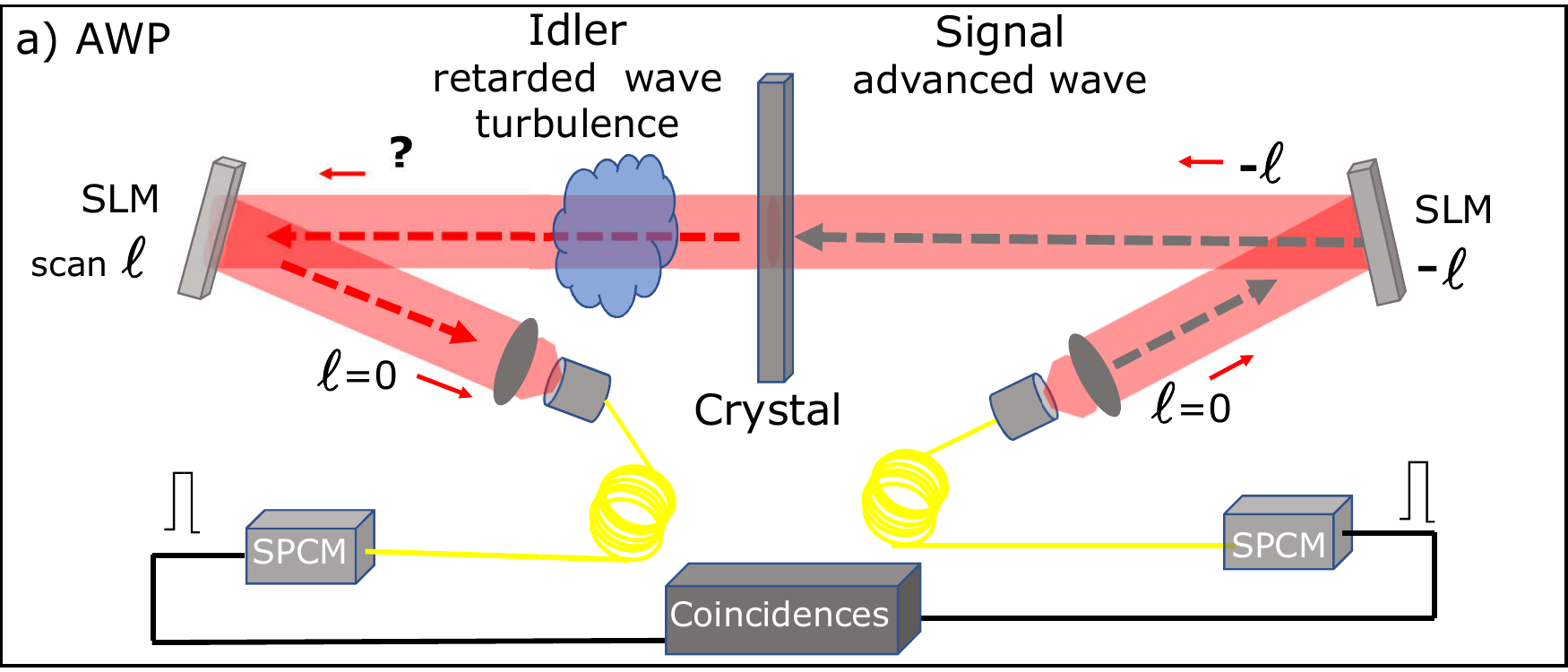}
\includegraphics[width=\columnwidth]{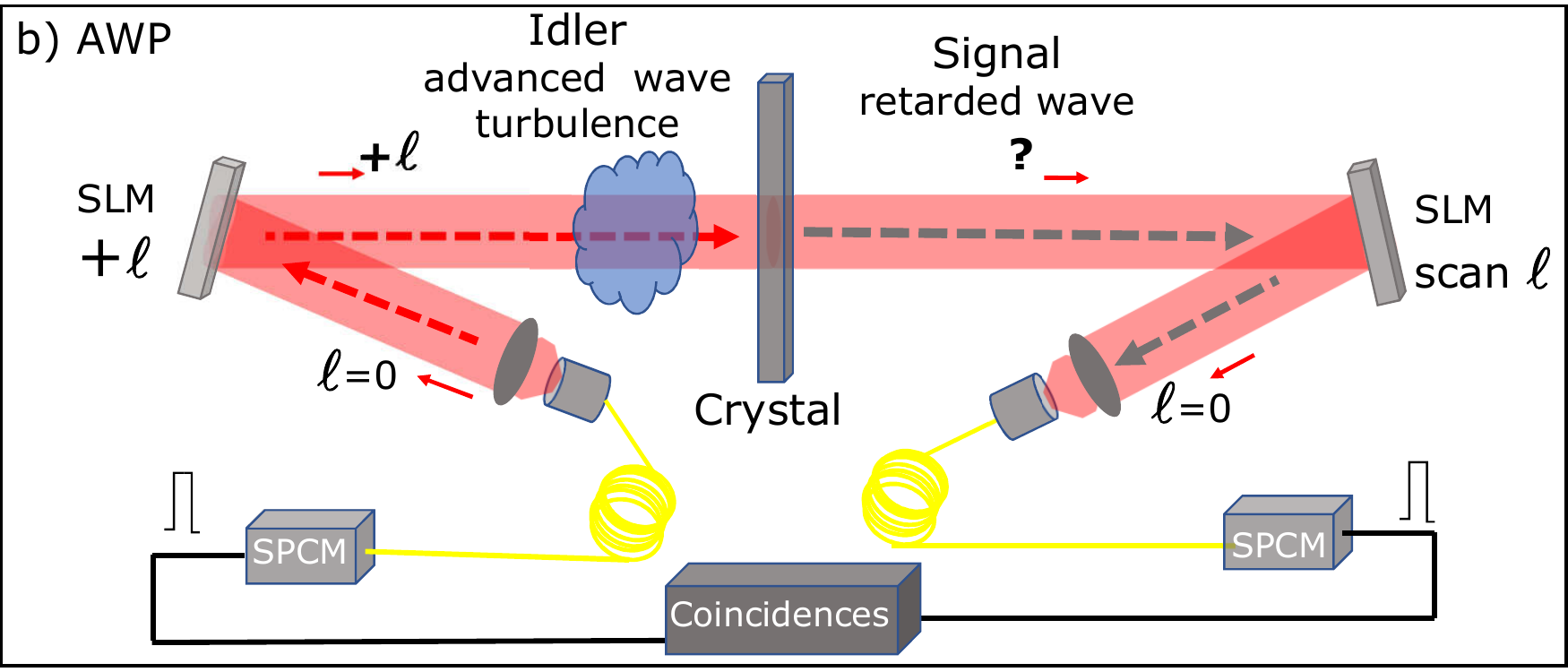}
\includegraphics[width=\columnwidth]{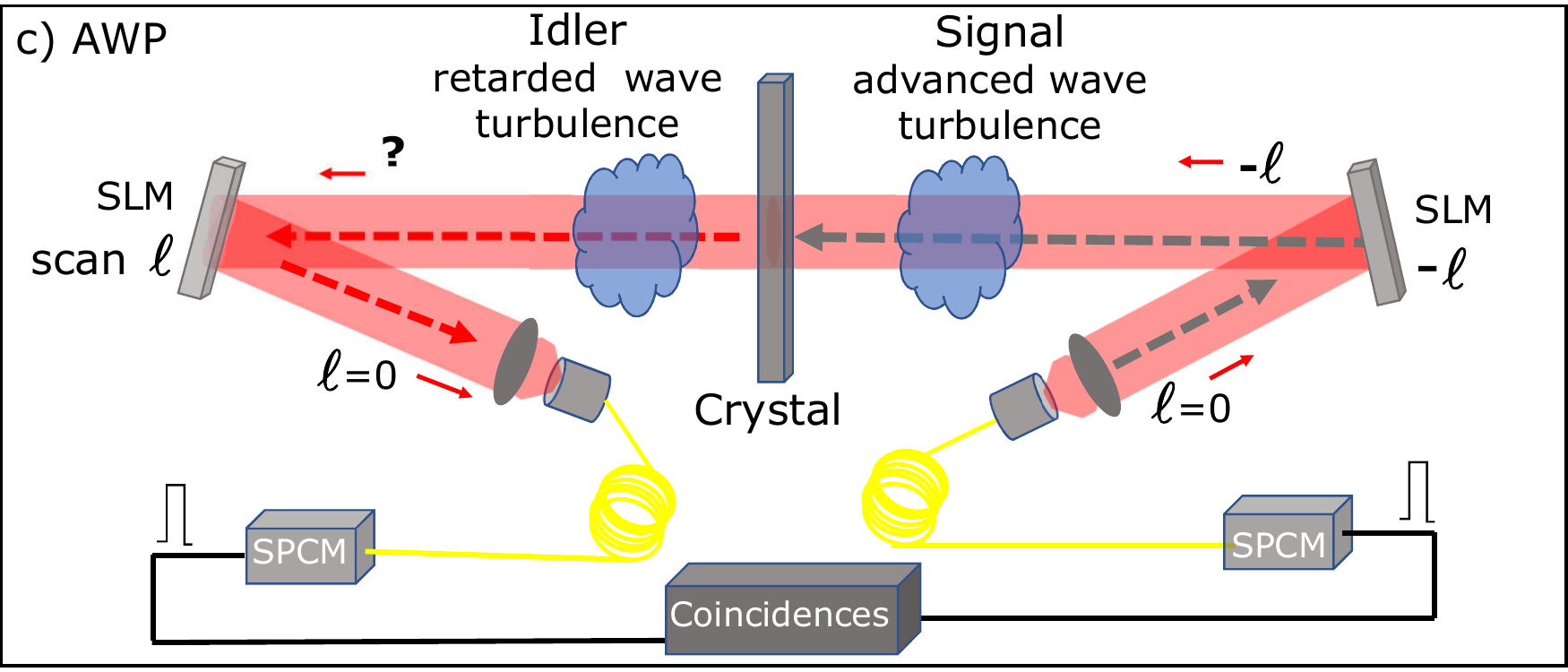}
\includegraphics[width=\columnwidth]{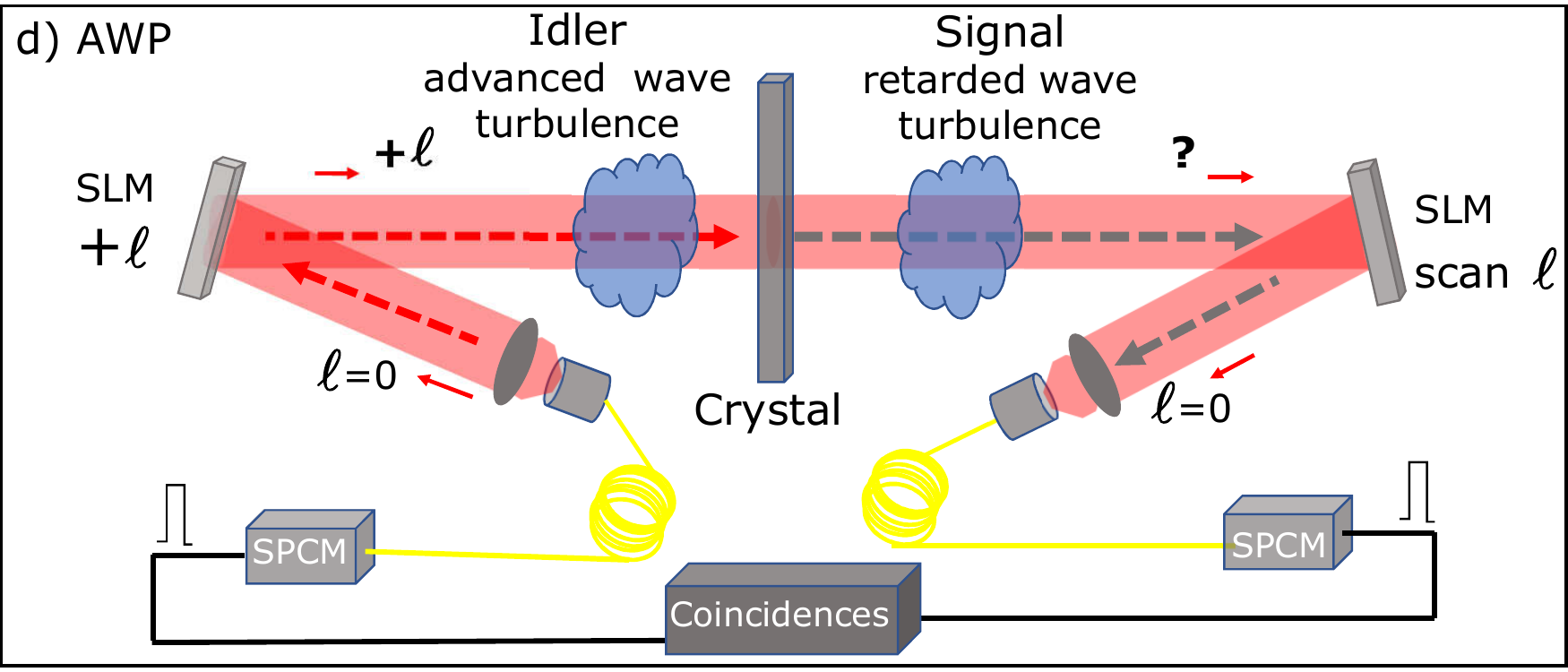}
\caption{Klyshko's advanced wave picture (AWP) of two-measurement protocol of a turbulent process. a) Direct and b) reverse AWP with single-photon turbulence. c) Direct and d) reverse AWP with two-photon turbulence. See main text for details.}
\label{AWP2}
\end{figure}

\vspace{0.2cm}
\noindent
{\bf The work protocol} --- Figure~\ref{AWP2}a) illustrates the AWP of the two-measurement protocol for a single-sided turbulence channel acting as a process on the idler photon. The advanced photon is emitted by the signal SPCM and comes out of the fiber with OAM = 0. It propagates towards the signal SLM, which applies $\cal{L}_{-}^{\ell}$, thus preparing a OAM state with index $-\ell$. Notice that the OAM for the signal retarded wave is $+\ell$, while the OAM for the advanced wave has the opposite sign, because they propagate in opposite senses and the OAM sign is defined depending on the sense of propagation using a right-hand rule. The crystal acts like a transparent window and the idler retarded wave is prepared with OAM = $-\hbar\ell$ per photon. This corresponds to the first measurement of the protocol, which actually prepares a pure energy eigenstate, represented here by a mode with given OAM. The photon propagates through the turbulence and couples to other OAM modes. The idler SLM is scanned, that is to say, measurements are made using a sequence of masks applying rising and lowering operators ranging from ${\cal L}_{-}^{m}$ to ${\cal L}_{+}^{m}$, where $\pm m$ are the maximal/minimal azimuthal indices. Coupling to the fiber and knowledge about the applied mask provides the measurement result. This corresponds to the second measurement of the protocol. From the outcome of these measurements we obtain the conditional probabilities that are necessary to compute the relevant quantites in Jarzynski's relation, as it will be explained in detail later.

Figure~\ref{AWP2}b) illustrates the backwards version of the process in Fig.~\ref{AWP2}a). Due to the symmetry in the AWP, it is also possible to interpret the two-measurement protocol supposing that the source of advanced waves is the idler SPCM and that the signal SPCM is the actual detector. Moreover, there are lenses (not shown in Figs. \ref{AWP1} and \ref{AWP2} for simplicity) that image input planes onto output planes (e.g. crystal plane is imaged onto both SLM planes), cancelling all free propagation effects. Therefore, AWP  symmetry between signal-to-idler and idler-to-signal senses is practically perfect. 

Figure~\ref{AWP2}c) shows a scheme similar to Fig.~\ref{AWP2}a), but using a two-sided turbulence channel. The interpretation in terms of AWP is identical to the case of single-sided channel. The main difference is that, now, both signal and idler photons are affected by turbulence, increasing the mode scattering. In other words, this process tends to couple the input OAM mode to a larger number of output OAM modes. Figure \ref{AWP2}d) shows the backwards version of Fig.~\ref{AWP2}c), again based on the AWP symmetry.

\vspace{0.5cm}
\appendix 

{\bf APPENDIX B}  \\  \\  {\it Loss of phase information in optical diffraction}\\ 

\begin{figure}[h]
\includegraphics[width=\columnwidth]{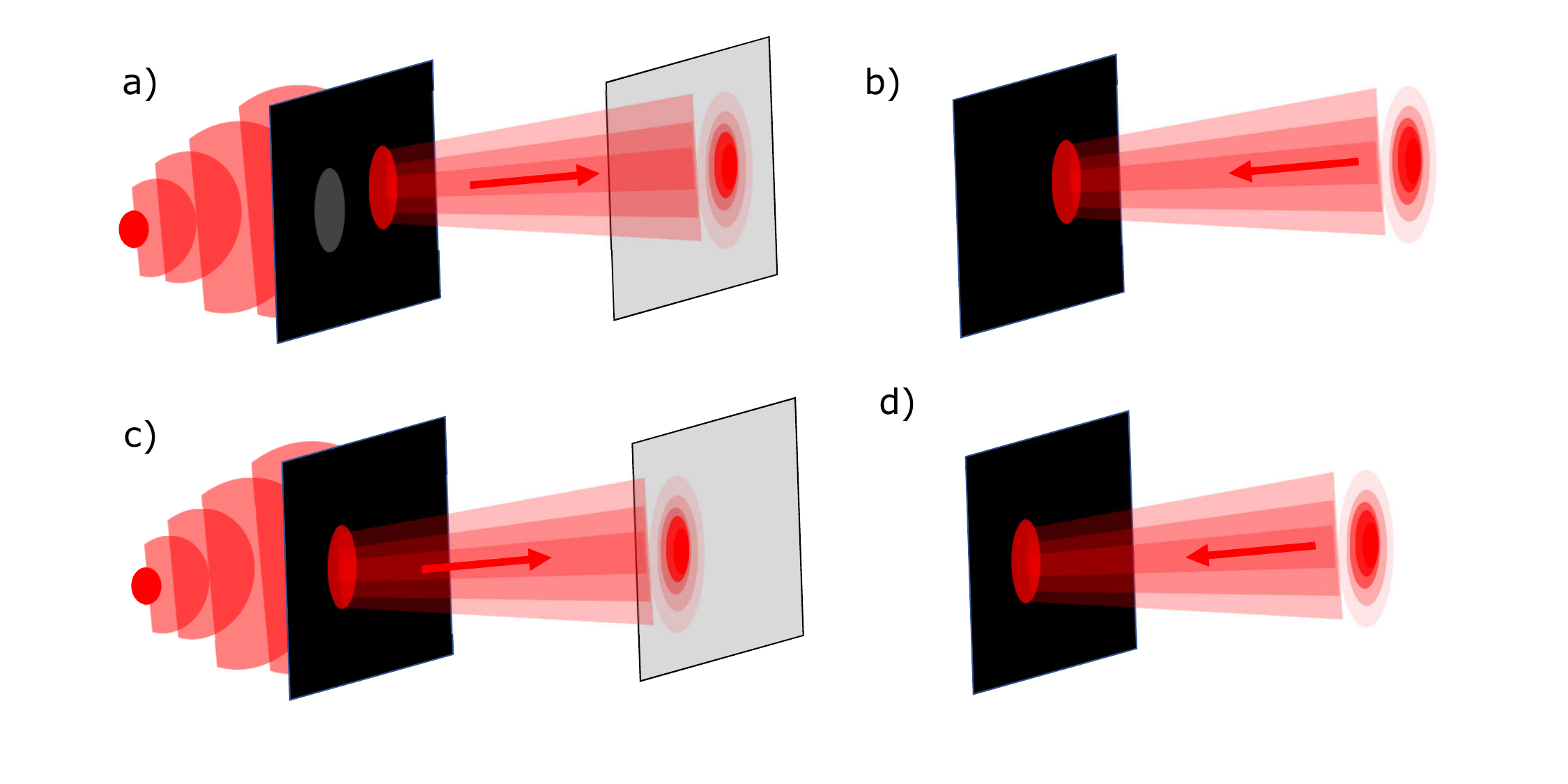}
\caption{a) Right single-slit diffraction and b) its time reversal. c) Left single-slit diffraction and d) its time reversal.}
\label{single}
\end{figure}
\begin{figure}[h]
\includegraphics[width=\columnwidth]{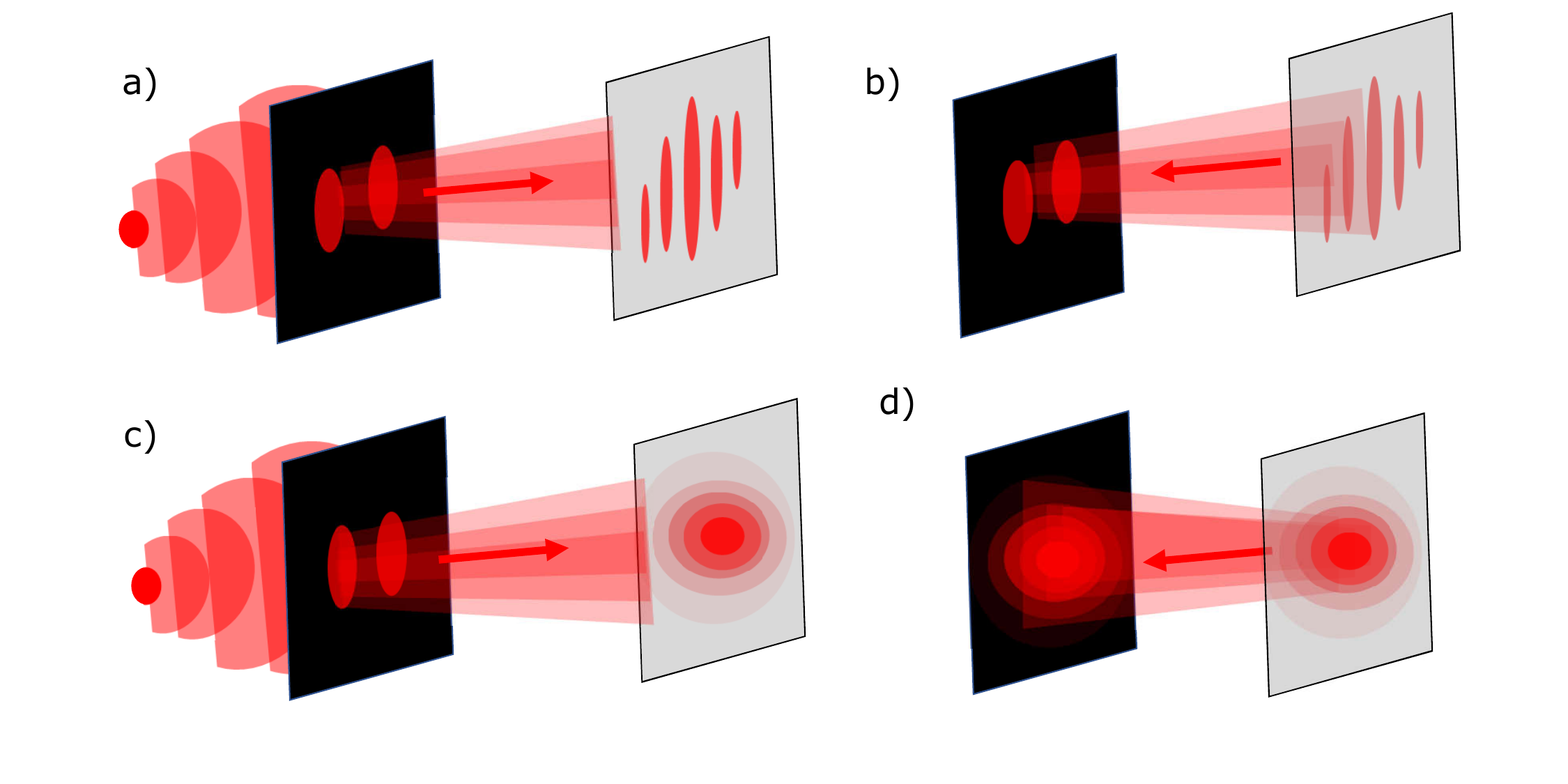}
\caption{a) Double slit diffraction from coherent sources; b) time reversal of the diffraction; c) double slit diffraction pattern for fluctuating phase differences; d) time reversal of pattern in c).}
\label{double}
\end{figure}

In our experiment we use entangled photon pairs that are subjected to diffraction in a phase mask generated by a spatial light modulator (SLM). The SLM masks simulate Komolgorov turbulence and in order to reproduce its dynamical effects, it is necessary to apply a few different masks with the same scintilation strength during the measurement time interval. This means that the results of a few diffraction patterns are summed uncoherently implying in loss of phase information and irreversibility. Even though we use coincidence detection approaches in our experiment the loss of phase information can be exaplained classically, because the propagation of signal and idler fields is made in the classical optics fashion. Please refer to Ref. \cite{Walborn10a} for details about the paraxial theory of spontaneous parametric down-conversion spatial correlations.

In this appendix, we illustrate the loss of phase information and irreversibility in optical diffraction in the context of the double-slit diffraction experiment.
In Fig. \ref{single}a) the single slit diffraction is illustrated. It is illuminated by a coherent source and produces a characteristic diffraction pattern. If one reverses the propagation of this field, it will come back to same initial distribution in the plane of the slit, as illustrated in Fig. \ref{single}b). It is worth mentioning that the time reversal of an optical field is physically feasible using a phase conjugation mirror \cite{Pepper82}. The same diffraction and time reversion happens for the other slit, as illustrated in Figs. \ref{single}c) and d).

In Fig. \ref{double}a) the double slit diffraction is illustrated. It is illuminated by a coherent source and produces a characteristic diffraction or interference pattern. The time reversal of the interference pattern returns to the same initial distribution, with two bright spots, as illustrated in Fig. \ref{double}b).  In Fig. \ref{double}c) the double slit is illuminated by an incoherent source. As a result, the phase difference between the field in the slits fluctuates and no interference pattern is observed. The intensity measurement by some device always make some time average of the intensity distribution. This is equivalent to measuring intensity patterns for several instants of time and summing up the result. After the intensity is measured, the phase information is lost and even if one tries to produce a new field with the same amplitude distribution and propagate it backwards as in Fig. \ref{double}d), the inital pattern would not be recovered without the appropriate phase information. 

\bibliographystyle{apsrev}

\end{document}